# Catalytic Synthesis of C2-C4-Alkenes from Dimethyl Ether

Shuxrat Chorievich Aslanov[1], Abdurazzoq Qobilovich Buxorov[2], Normurot Ibodullaevich Fayzullayev[3*]

*[1] CEO, Shurtan gas chemical complex, Kashkadarya, Uzbekistan*
ORCiD: https://orcid.org/0000-0003-0179-364X

*[2] Head of the training center, Shurtan gas chemical complex, Kashkadarya, Uzbekistan*
ORCiD: https://orcid.org/0000-0002-7765-3152

*[3] Doctor of Technical Sciences, Professor, Department of Polymer Chemistry and Chemical Technology, Samarkand State University, 140104, University Blv. 15, Samarkand, Uzbekistan.*
ORCiD: https://orcid.org/0000-0001-5838-3743

[1*]f-normurot@samdu.uz, [2]aslanov@sgcc.uz; [3]buxorov@mail.ru,

*Abstract -* *In the study, the effect of the volumetric velocity of the initial gaseous mixture on the catalytic properties of Zn-Zr-Cu\*HSZ\*Al2O3 in the conversion of dimethyl ether to olefins was studied. When the volumetric rate is increased by 5 times, the conversion of dimethyl ether decreases from 96.5 to 40.7%, the total selectivity for $C_2 = - C_4 =$ olefins decreases from 59.0 to 75.5 mas.%, at the same time, the yield of the by-products of the reaction - $C_2$ + paraffin is reduced. A decrease in the reaction temperature at the equal conversion of dimethyl ether leads to an increase in selectivity for both ethylene and propylene due to a decrease in the formation of methane and $C_2$ + paraffin in the reaction products. Thus, in copper-containing zeolite catalysts, lower olefins than dimethyl ether provide higher conversion of raw material and higher selectivity of olefin formation (up to 90 mas. %). Also, changing the initial raw material composition and regime parameters allows a significant change in the ratio of olefins to each other. To study the primary intermediates of the formation of olefins from dimethyl ether, i.e., the conversion of the C-O bond to the primary C-C bond, the copper-storage catalyst was tested at atmospheric pressure and dimethyl ether conversion at 240 °C for the homologation reaction of methanol. The data set obtained to the prospects of the catalytic system Zn-Zr-Cu\*HSZ\*Al2O3 modified with copper compounds. The selectivity for olefins reaches 87-90%.*

## I. INTRODUCTION

A new three-step method of deep processing of natural gas into lower olefins by synthesis-gas (but in the direction of its conversion to dimethyl ether is carried out as "methanol" processes with subsequent conversion of DME to ethylene and propylene) has been developed:

Methane → synthesis gas → DME → lower olefins

The conversion of methane to ethylene by dimethyl ether has several advantages over the methanol method. Hence, more favorable thermodynamics allow the synthesis of DME at lower temperatures than the synthesis of methanol and achieve a much deeper conversion of the synthesis gas in a single pass through the catalyst [1]. As a result, energy and material costs are reduced, as well as the efficiency of natural gas use increases, and the cost of production decreases [2]. Other advantages of this method:

- Decrease in thermal stress due to the removal of the dehydrated exothermic reaction of methanol from this stage during the synthesis of olefins from DME [3];

- Much higher activity and selectivity of DME in the synthesis of olefins compared with methanol;

- Along with ethylene and propylene, DME is a new valuable product available in the chemical and fuel markets, the demand for which has increased more than 400 times since the beginning of the century [4].

Dimethyl ether is considered as an intermediate raw material in the production of alternative environmentally friendly municipal and transport fuels, fuels for gas turbines, synthetic high-octane gasoline, polymer products, and organic synthesis products [5-12]. Several large industrial devices for the production of DME from synthetic gas are operating in China, Japan, Russia, the Far East. The rounding of production of olefins (polyolefins) allows expanding the range of products produced in these industrial schemes, where the direct synthesis of DME from synthesis gas reduces production risks and creates an economic safety cushion for complex multi-tonnage production [13-18].

## II. EXPERIMENTAL PART

The high-modulus zeolite used to prepare the catalyst is a local analog of foreign zeolite. The hydrogenated form of zeolite (HHSZ) was obtained by drying after double cation exchange in ammonium nitrate solution with Na+ 1 N with a fixed residual amount of sodium oxide in it and annealing at 500 °C for 4 hours [19-22]. Zeolite-containing HSZ catalysts were obtained by mixing the zeolite with a binder (a suspension of aluminum oxide) and subsequent molding into granules (extruders). The extruders were air-dried, then oven-dried and burned at 500 °C for 4 hours.

The active elements to the zeolite were introduced into the initial zeolite by the method of adsorption (Zn, Zr) until mixing with the metal Tuzi aqueous solution of the binders (Zn, ZR), by the method of adsorption of ready-made zeolite extruders (Cu, Mg) together with the binders (Cu, Mg), or by the method of ion exchange (Mg).

Dimethyl ether was used as an initial raw material, He, $N_2$ or CO + $H_2$ as the diluent. The experiments were performed on a laboratory device using a flow-type microreactor. 0.3–1.0 g (0.6–2.0 ml) of catalyst (0.2–1.6 mm fraction) was added to the quartz reactor. Then, the catalyst activation was carried out at 400 °C in the *He* stream. The required volumetric velocity (500–1000 $h^{-1}$), temperature (240–450 °C), and pressure (~ 1 atm) of the reagents were determined. The condensed liquid products were collected in a collector, and the gas stream was sent for chromatographic analysis using a tap-dispenser (loop volume 1 ml).

The reaction gas stream of dimethyl ether conversion consists of a mixture of hydrocarbons $C_1$-$C_6$. The main method of their analysis is gas-liquid chromatography (GLC). Analysis of gas impurities was carried out on a Crystal lux 4000M chromatograph with a flame-ionization detector. The dimensions of the capillary column 27.5 m*0.32 mm (Varian, USA), as an adsorbent, the non-polar phase of CP-paraplot was used (thickness of the adsorption layer 10 μm), which was effective enough to separate the main group of phase reaction products (dimethyl ether, $CH_3OH$, $C_1$-$C_6$ hydrocarbons). The analysis was carried out in a thermo-programmed mode (80-150 °C, heating rate 30 °C/min), carrier-gas - helium (flow separation 1:8). The resulting chromatograms were processed by a NetchromWin program.

## III. RESULTS AND DISCUSSION

It is known that high-modular pentacyl, an analog of HSZ, is an effective catalyst for the conversion of dimethyl ether to lower olefins ($C_2$-$C_4$). However, it has insufficient selectivity on these valuable products.

The modified zeolite catalyst with Zn and Zr is compared with the hydrogen form HSZ allows to increase selectivity up to 80 mas.% on lower olefins at ~ 70% conversion of dimethyl ether. The ratio of ethylene to propylene is ~ 1.1. Further improvement of catalytic properties can be ensured by selecting the optimal modifying element and changing the technological parameters of the process.

To solve this problem and implement it in industry, it is necessary to study the intermediates that lead to the formation of ethylene as the primary product of the reaction, as well as the involvement of oxygenates formed in parallel with this reaction. The effect of modification of zeolite catalysts with copper compounds has been studied. They are well active components of the carbonization and homologation reactions of methanol. Based on preliminary experiments, it was determined that it is expedient to introduce copper by the method of absorption of finished extruders HSZ*Al₂O₃.

The effect of copper concentration on the zeolite catalyst on its catalytic properties in dimethyl ether conversion was studied. The data in Table 1 show that under these conditions, the reaction proceeds with almost complete conversion of dimethyl ether; selectivity for ethylene by 4.4 mas.% with an increase in copper concentration from 0.05 to 0.1 mas.%, selectivity for propylene increases by 1.1 mas.%. Subsequent increases in copper concentrations are observed with a decrease in selectivity for ethylene and a slight increase in selectivity for propylene. The best results were obtained at a concentration of 0.1 mas.% of copper.

**TABLE I.  INFLUENCE OF THE CONCENTRATION OF COPPER IN THE ZEOLITE CATALYST Cu\*HSZ\*Al2O₃ ON ITS CATALYTIC PROPERTIES DURING THE CONVERSION OF DIMETHYL ETHER TO OLEFINS**

| [Cu], мас.% | $K_{DME}$, % | Selectivity, mas.% | | | | | | |
|---|---|---|---|---|---|---|---|---|
| | | $CH_4$ | $C_{2=}$ | $C_{3=}$ | $C_{4=}$ | $C_{5=}$ | $\sum C_{2+}$ | $\sum C_{2=} - C_{4=}$ |
| 0,05 | 99,6 | 0,8 | 19,9 | 18,3 | 7,9 | 3,5 | 50,4 | 46,1 |
| 0,1 | 99,9 | 0,2 | 24,3 | 19,4 | 6,3 | 2,7 | 47,1 | 50,0 |
| 0,2 | 99,5 | 0,9 | 21,4 | 20,3 | 7,1 | 3,0 | 48,2 | 48,8 |

(T = 340 °C, P = 0,1 MPa; initial mixture: 10% dimethylether + He, $V_{mix} = 1000$ $h^{-1}$)

The effect of the nature of copper compounds on the catalytic properties of the zeolite catalyst has been studied. From the data in Table 2, the modified sample with copper complex (Polyethylenimine - PEE) showed the greatest selectivity (50.0-75.2 mas. %) for $C_2$-$C_4$ olefins.

**TABLE II. INFLUENCE OF THE NATURE OF COPPER COMPOUNDS ON THE CATALYTIC PROPERTIES OF THE ZEOLITE CATALYST**

| [Cu] compound | T, °C | $K_{DME}$, % | Selectivity, mas.% | | | | | | |
|---|---|---|---|---|---|---|---|---|---|
| | | | $CH_4$ | $C_{2=}$ | $C_{3=}$ | $C_{4=}$ | $C_{5=}$ | $\sum C_{2+}$ | $\sum C_{2=} - C_{4=}$ |
| $CuCl_2$ | 320 | 79,1 | 1,3 | 29,2 | 27,2 | 14,6 | 3,8 | 23,9 | 71,0 |
| | 340 | 99,7 | 1,2 | 18,6 | 11,4 | 10,1 | 5,4 | 53,3 | 40,1 |
| AcAc | 320 | 97,2 | 1,6 | 24,5 | 14,3 | 15,2 | 6,1 | 38,3 | 54,0 |
| | 340 | 99,6 | 1,6 | 10,9 | 7,4 | 8,7 | 3,4 | 68,0 | 27,0 |
| PEE | 320 | 87,4 | 0,4 | 28,2 | 35,4 | 11,6 | 0,9 | 23,5 | 75,2 |
| | 340 | 99,9 | 0,2 | 24,3 | 19,4 | 6,3 | 2,7 | 47,1 | 50,0 |

To study the primary intermediates in the formation of olefins from dimethyl ether, i.e., the conversion of the C-O bond to the primary C-C bond, the copper-storage catalyst was tested at atmospheric pressure and dimethyl ether conversion at 240 °C for the homologation reaction of methanol. As can be seen from the data in Table 3, significant amounts of methanol (up to 44 mas. %) and up to 8.2 ms.% of ethanol at the sufficiently low conversion of dimethyl ether to reaction products under these conditions, as well as large amounts of alkanes at the sufficiently low conversion of dimethyl ether were detected. As the reaction temperature increased to 270 °C, the selectivity for ethylene and propylene increased significantly. The amount of ethanol and methanol decreased. In particular, the amount of ethanol was significantly reduced. The methanol/ethanol ratio increased from 5.4 to 12.6. Cu*HSZ*Al₂O₃

**TABLE III. DIMETHYL ETHER CONVERSION IN Cu(0,1*HSZ*Al₂O₃ CATALYST**

| T, °C | $K_{DME}$, % | Selectivity, mas.% | | | | | | | |
|---|---|---|---|---|---|---|---|---|---|
| | | $CH_4$ | $C_{2=}$ | $C_{3=}$ | $C_{4=}$ | $C_{5=}$ | $\sum C_{2+}$ | Methanol | Ethanol |
| 240 | 8,6 | 1,0 | 5,8 | 16,9 | 0,1 | - | 24,1 | 43,9 | 8,2 |
| 270 | 20,4 | 1,5 | 22,0 | 29,2 | 0,1 | 2,3 | 7,1 | 35,2 | 2,8 |

The study of the effect of volumetric velocity on the ratio of methanol/ethanol in reaction products (see Figure 1) shows that with an increase in volumetric speed, the mass ratio of methanol/ethanol increases, which is based on the assumption that methanol is an intermediate product in the formation of ethanol.

To test this assumption, methanol experiments were performed in a synthesis gas atmosphere to determine the contribution of methanol homologation (see Table 4). However, surprisingly, at temperature 240 °C, the conversion of methanol in the catalyst Cu*HSZ*Al₂O₃ takes place at a very low temperature, and the main component in the products is dimethyl ether. Ethanol in the reaction products is also three-fold, but its content is significantly lower compared to its content in dimethyl ether products under the same conditions (Table 3). Based on the results presented, it can be assumed that the contribution of methanol to ethanol formation in the homologous reaction is significantly smaller compared to the contribution of the dimethyl ether isomerization reaction.

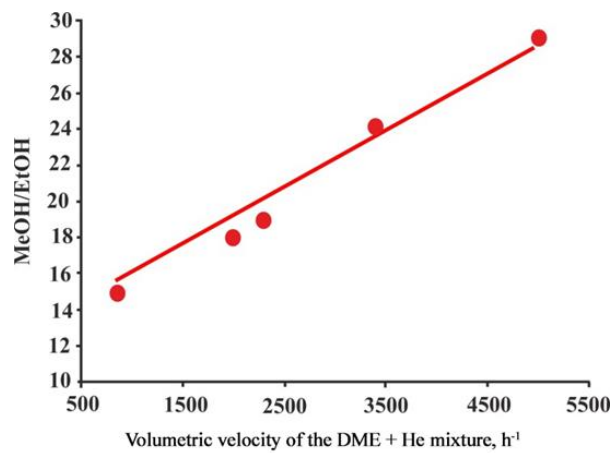

**Figure 1. Dependence of the mass ratio of methanol/ethanol on Cu (0.1)*HSZ*Al₂O₃ on the volumetric velocity of the initial gaseous mixture (T=270 °C, initial mixture: 10% dimethyl ether +He).**

**TABLE IV.  METHANOL CONVERSION IN Cu (0,1)* HSZ*Al2O3 CATALYST**

| T, ℃ | $K_{DME}$, % | Selectivity, mas.% | | | | | | | |
|---|---|---|---|---|---|---|---|---|---|
| | | CH4 | C2= | C3= | C4= | C5= | $\sum C_{2+}$ | Methanol | Ethanol |
| 240 | 9,4 | 1,2 | 2,7 | 2,8 | - | - | 2,1 | 90,6 | 0,6 |
| 270 | 78,3 | 11,9 | 8,2 | 16,3 | - | 0,4 | 37,9 | 21,3 | 4,0 |

The conversion of dimethyl ether at low temperatures in a zinc-zirconium catalyst modified with a copper complex (Zn-Zr-Cu(0,1)*HSZ*Al2O3) is accompanied by significantly lower production of ethanol than in Cu(0,1)*ЮКЦ*Al2O3. The conversion of helium to a synthesis gas resulted in a slight increase in dimethyl ether conversion and an increase in the overall selectivity of $C_2$- $C_{4=}$ for olefins, with more propylene growth.

**TABLE V. CONVERSION OF DIMETHYL ETHER AND METHANOL IN Zn-Zr-Cu(0,1)* HSZ*Al2O3**

| Initial compound | T, ℃ | $K_{ДМЭ}$, % | Selectivity, mas.% | | | | | | |
|---|---|---|---|---|---|---|---|---|---|
| | | | CH4 | $C_{2=}$ | $C_{3=}$ | $C_{4=}$ | $\sum C_{2+}$ | Methanol | Ethanol |
| Dimethyl ether +He | 240 | 0,5 | 0,9 | 4,6 | 14,2 | 0,5 | 0,8 | 79,0 | - |
| | 270 | 1,7 | 0,4 | 7,1 | 30,8 | 1,8 | 1,3 | 57,1 | 1,5 |
| Dimethyl ether +CO+H | 240 | 0,6 | 0,9 | 8,8 | 28,4 | - | 3,8 | 56,9 | 1,2 |

With the increase in temperature to 340 ℃, the selectivity for the formation of olefins from dimethyl ether in copper-modified samples increased sharply (Table 6). Hence, the dimethyl ether conversion increases from 20.4 to 100% for the Cu (0.1)*HSZ*Al2O3 catalyst as the temperature increases from 270 ℃ to 340 ℃. The selectivity for $\sum C_{2=} - C_{4=}$ olefins are maintained almost unchanged. In contrast, dimethyl ether conversion in Zn-Zr-Cu(0,1)*HSZ*Al2O3 increased by only 73%, while selectivity for $\sum C_{2=} - C_{4=}$ olefins increased from 39.7 to 87 mas.%. In the case of methanol, the olefin selectivity at 100% conversion was 57.3 mas. %. Carrying out the dimethyl ether change reaction in the synthesis-gas stream leads to a sharp increase in conversion and a change in the composition of olefins towards an increase in propylene of 90 mas.% in the almost unchanged total yield of olefins $C_{2=}$- $C_{4=}$. Selectivity for olefins reaches 87-90%.

**TABLE VI.  CONVERSION OF DIMETHYL ETHER AND METHANOL UNDER THE CONDITIONS OF SYNTHESIS OF OLEFINS**

| Catalyst | Initial compound | $K_{DME}$, % | Selectivity, mas.% | | | | | | |
|---|---|---|---|---|---|---|---|---|---|
| | | | CH4 | $C_{2=}$ | $C_{3=}$ | $C_{4=}$ | $C_{5=}$ | $\sum C_{2+}$ | $\sum C_{2=} - C_{4=}$ |
| Cu*HSZ*Al2O3 | DME (10%)+He | 99,9 | 0,2 | 24,3 | 19,4 | 6,3 | 2,7 | 47,1 | 50,0 |
| Zn-Zr*HSZ*Al2O3 | DME (10%)+He | 73,0 | 1,1 | 28,1 | 36,6 | 6,2 | 2,2 | 25,8 | 70,9 |
| Zn-Zr-Cu*HSZ*Al2O3 | DME (10%)+He | 73,0 | 0,2 | 30,4 | 37,9 | 18,7 | 0,8 | 12,0 | 87,0 |
| | CH3OH(10%)+CO(30%)+H2(60%) | 100 | 2,4 | 28,1 | 22,2 | 7,0 | - | 40,3 | 57,3 |
| | DME(10%)+CO(30%)+H2(60%) | 96,0 | 0,5 | 32,0 | 48,0 | 10,1 | 0,9 | 8,5 | 90,1 |

Based on the results obtained, a scheme was proposed for the synthesis of lower olefins from dimethyl ether, in which one of the main intermediates is ethanol, which can be formed both during the isomerization of dimethyl ether and during the homologation of methanol (fig. 2.).

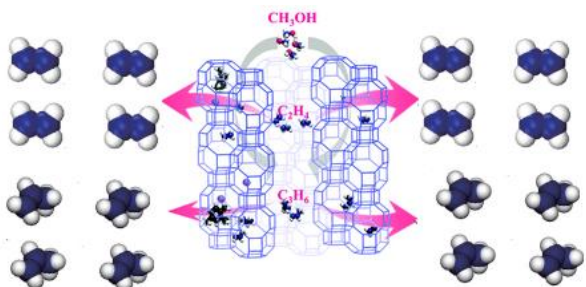

**Figure 2. Formation of ethylene and propylene from dimethyl ether or methanol**

Taking into account the prospects of the developed catalytic system, the effect of the volumetric velocity of the initial gas mixture on the catalytic properties of Zn-Zr-Cu*HSZ*Al₂O₃ during the conversion of dimethyl ether to olefins has been studied. With a 5-fold increase in space velocity, the conversion of dimethyl ether decreases from 96.5 to 40.7%, and the overall selectivity for olefins $C_{2=}$- $C_{4=}$ decreases from 59.0 to 75.5 mas.%. At the same time, the yield of reaction by-products, $C_2$ + paraffin, decreases.

A decrease in the reaction temperature at the equal conversion of dimethyl ether leads to an increase in selectivity for both ethylene and propylene due to a decrease in the formation of methane and $C_2$ + paraffin in the reaction products.

Thus, in copper-containing zeolite catalysts, lower olefins than dimethyl ether provide higher conversion of raw material and higher selectivity of olefin formation (up to 90 mas.%). Also, changing the initial raw material composition and regime parameters allows a significant change in the ratio of olefins to each other. The primary zeolite-preserving catalyst (HSZ*Al₂O₃) is modified with magnesium compounds, it is known from the literature that it can interact effectively with alkyl fragments and perform magnesium synthesis, and it leads to isomerization of hydrocarbons with functional groups.

Table 7 shows the catalytic properties of samples prepared by different methods of magnesium introduction. Dimethyl ether conversion changed slightly when changing the magnesium injection method. However, the ratio of the components of the lower olefins and their total yield was significantly dependent on the route of introduction of magnesium into the zeolite catalyst. The ethylene/propylene ratio changed from 1.9 to 1.1, i.e., 2 times. In practice, it is possible to change the production of ethylene, propylene, and butylene depending on the factory requirements in real conditions. Good results in terms of the total yield of lower olefins were obtained in the introduction of magnesium by ingestion of HSZ*Al₂O₃ ready-made extruders. At the same time, propylene yields increased significantly. The lowest yield of methane is observed for the same sample.

**TABLE VII. INFLUENCE OF METHODS OF INTRODUCTION OF MAGNESIUM ON A ZEOLITE CATALYST ON ITS CATALYTIC PROPERTIES**

| Magnesium addition method | $K_{DME}$, % | Selectivity, mas.% | | | | | | | |
|---|---|---|---|---|---|---|---|---|---|
| | | $CH_4$ | $C_{2=}$ | $C_{3=}$ | $C_{4=}$ | $C_{5=}$ | $\sum C_{2=} - C_{4=}$ | $\sum C_{2+}$ | $C_{2=}*C_{3=}$ |
| | 69,2 | 2,7 | 29,4 | 15,2 | 4,6 | 0,1 | 49,2 | 48,0 | 1,9 |
| | 67,4 | 2,2 | 30,8 | 19,2 | 6,3 | 0,2 | 56,3 | 41,3 | 1,6 |
| | 70,5 | 1,9 | 26,7 | 15,3 | 11,0 | 0,1 | 53,0 | 45,0 | 1,7 |
| | 69,3 | 0,7 | 33,1 | 29,3 | 1,2 | 0,8 | 63,6 | 34,9 | 1,1 |

Based on the results obtained, further studies were carried out with samples prepared by absorption from finished catalyst extruders, if not only the zeolite component but also the alumina binder were modified with magnesium. The results shown in Table 8 show that the conversion of dimethyl ether decreases as the amount of magnesium introduced into the catalyst increases, while the selectivity of olefin formation $C_2$-$C_4$ passes through the maximum. The best results were obtained on a catalyst containing 1.0 wt% Magnesium. The total selectivity for $C_2$-$C_4$ olefins was about 80.8 mas.%. Almost 80% of them are ethylene and propylene. Magnesium-preserving samples high yields of butylene were observed at 360 °C. (up to 14.6 mas.%). This can be of practical importance. The ratio of ethylene/propylene was slightly reduced when the amount of magnesium included was increased. The total yield of alkanes decreased significantly when the magnesium content was increased to 1.0% and remained almost unchanged when the magnesium content in the catalyst was subsequently increased. Very low coke formation is characteristic for all samples of magnesium-containing zeolite catalyst.

**TABLE VIII. INVESTIGATION OF THE EFFECT OF TESTS ON THE CATALYTIC PROPERTIES OF ZEOLITE CATALYST AT DIFFERENT TEMPERATURES**

| [Mg], mas.% | T, °C | $K_{DME}$, % | Selectivity, mas.% | | | | | | | |
|---|---|---|---|---|---|---|---|---|---|---|
| | | | $CH_4$ | $C_{2=}$ | $C_{3=}$ | $C_{4=}$ | $C_{5=}$ | $\sum C_{2+}$ | $\sum C_{2=} - C_{4=}$ | $C_{2=} * C_{3=}$ |
| 0,05 | 320 | 69,3 | 0,7 | 33,1 | 29,3 | 1,2 | 0,8 | 63,6 | 34,9 | 1,1 |
| | 340 | 96,8 | 0,7 | 27,4 | 16,2 | 10,4 | 0,9 | 54,0 | 44,4 | 1,7 |
| | 360 | 97,4 | 0,7 | 17,3 | 18,4 | 11,9 | 2,1 | 47,6 | 49,6 | 0,9 |
| 0,1 | 320 | 661,3 | сл. | 40,7 | 38,5 | 1,6 | 0,9 | 80,8 | 18,3 | 1,1 |
| | 340 | 85,6 | сл. | 33,6 | 21,6 | 12,0 | 2,4 | 67,2 | 30,4 | 1,6 |
| | 360 | 86,1 | сл. | 21,3 | 24,2 | 14,6 | 2,9 | 60,1 | 37,0 | 0,9 |
| 0,2 | 320 | 53,2 | сл. | 39,0 | 40,1 | 0,1 | 0,4 | 79,2 | 20,4 | 1,0 |
| | 340 | 74,3 | сл. | 32,2 | 22,2 | 11,1 | 2,4 | 65,5 | 32,1 | 1,5 |
| | 360 | 74,7 | сл. | 20,4 | 25,2 | 12,7 | 3,2 | 58,3 | 38,5 | 0,8 |

When the temperature of the reaction is increased from 320 °C to 360°C, the simultaneous decrease in the selectivity of ethylene and propylene formation indicates an increase in the selectivity of butylene formation and the sum of alkanes in the data obtained, indicating that the activation energy of ethylene and propylene formation reactions is lower than the activation energies of alkane formation and olefins polymerization reactions. It has been found that changing the concentration of dimethyl ether to 30% has a significant effect on the activity and selectivity of the catalyst. Given the potential prospects of a magnesium-containing zeolite catalyst, the effects of some technological parameters on dimethyl ether conversion and selectivity on lower olefins have been studied. To assess the potential effect of external diffusion inhibition in the conversion reaction of dimethyl ether to olefins, experiments were carried out at different linear velocities of the raw material flow in the reactor at a constant volumetric rate of feed delivery per unit time when calculating the catalyst volume. The catalytic properties of the catalyst samples were studied with particles of different sizes to evaluate the potential effect of external diffusion braking. The experiments were performed at a maximum possible temperature for this reaction - 450 °C, as the external and internal diffusion braking increases with increasing reaction temperature. As can be seen from the data in Tables 9–10, even at the highest temperature of the reaction, external and internal diffusion braking are not noticeable, and therefore it is safe to assume that the reaction takes place in the kinetic field in the studied range of temperatures. In subsequent experiments, the variable parameters of the process were changed according to the temperature of the reaction from 320 °C to 450 °C and the volumetric rate of flow of the raw material from 500 to 10000 $h^{-1}$.

**TABLE IX. EFFECT OF THE LINEAR VELOCITY OF THE INITIAL GAS MIXTURE ON DIMETHYL ETHER CONVERSION AND DISTRIBUTION OF REACTION PRODUCTS**

| Catalyst m(g)*V(ml) | Linear speed, m/c | t, hour | $K_{DME}$, % | Selectivity, mas.% | | | | | |
|---|---|---|---|---|---|---|---|---|---|
| | | | | $CH_4$ | $C_{2=}$ | $C_{3=}$ | $C_{4=}$ | $C_{5=}$ | $\sum C_{2+}$ |
| 1,0/2,0 | 2,5*10$^{-2}$ | 1 | 100 | 0,7 | 15,5 | 30,6 | 14,5 | 4,6 | 34,1 |
| | | 5 | 100 | 0,7 | 15,2 | 32,4 | 15,8 | 4,8 | 31,1 |
| 0,5/1,0 | 1,2*10$^{-2}$ | 1 | 100 | 0,6 | 14,5 | 29,0 | 16,0 | 7,1 | 32,8 |
| | | 5 | 100 | 0,7 | 14,7 | 31,7 | 18,2 | 4,9 | 29,9 |
| 0,3/0,6 | 7,4*10$^{-2}$ | 1 | 100 | 0,6 | 15,1 | 30,7 | 15,3 | 4,5 | 33,7 |
| | | 5 | 100 | 0,7 | 15,1 | 32,3 | 17,1 | 4,7 | 30,0 |

**TABLE X.  EFFECT OF CATALYST PARTICLE SIZE ON DIMETHYL ETHER CONVERSION AND DISTRIBUTION OF REACTION PRODUCTS**

| Catalyst fraction, mm | Catalyst m(g)*V(ml) | t, hour | $K_{DME}$, % | Selectivity, mas.% | | | | | |
|---|---|---|---|---|---|---|---|---|---|
| | | | | $CH_4$ | $C_{2=}$ | $C_{3=}$ | $C_{4=}$ | $C_{5=}$ | $\sum c_{2+}$ |
| 0,2-0,4 | 0,5*1,0 | 1 | 100 | 0,7 | 14,8 | 35,5 | 17,4 | 5,1 | 26,5 |
| | | 5 | 100 | 1,0 | 14,8 | 35,4 | 17,9 | 5,8 | 25,1 |
| 0,4-0,6 | 0,5*1,0 | 1 | 100 | 0,6 | 14,5 | 29,0 | 16,0 | 7,1 | 32,8 |
| | | 5 | 100 | 0,7 | 14,7 | 31,7 | 18,2 | 4,9 | 29,9 |
| 0,6-1,0 | 0,5*1,05 | 1 | 100 | 0,6 | 15,7 | 31,9 | 14,9 | 4,7 | 32,1 |
| | | 5 | 100 | 0,7 | 15,4 | 33,9 | 16,7 | 5,1 | 28,2 |
| 1,0-1,6 | 0,5*1,06 | 1 | 100 | 0,7 | 13,9 | 31,6 | 16,8 | 4,7 | 32,3 |
| | | 5 | 100 | 0,6 | 15,1 | 34,3 | 16,8 | 5,5 | 27,8 |

Figure 2 and Table 11 show the data on the selectivity of product formation for the limiting values of temperatures (450 °C and 320 °C) at different volumetric velocities of raw material delivery. At 450 °C, the rate of conversion of dimethyl ether is so high that the conversion in the whole studied range of volumetric velocities of raw material delivery is 100 mas. %. However, as can be seen from Figure 3, increasing the mass rate of raw material delivery at 100% conversion of dimethyl ether leads to a significant change in the composition of the products resulting from changes in the transition conditions of the secondary reactions. A decrease in contact time (increase in mass velocity) will likely lead to a decrease in the contribution of secondary reactions. In almost such a decrease in ethylene yield, a flat increase in propylene yield was observed, indicating a decrease in propylene conversion secondary reactions. At the maximum mass rate of delivery of raw materials in the liquid for 6 hours-1, the propylene/ethylene ratio is 3.1 and varies in the direction of the regime carried out by Lurgi in the process of "methanol to propylene." The yield relationship between isobutane and ethylene, on the one hand, and isobutane and N-butene with ethylene, on the other hand, also draws attention.

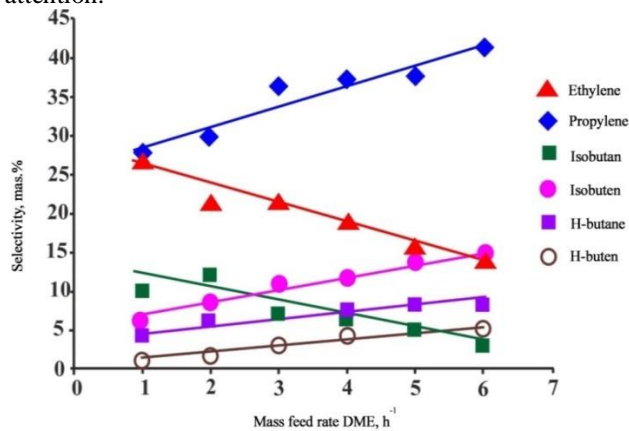

**Figure 3. The effect of the mass rate of dimethyl ether delivery on the distribution of reaction products**

(Dimethyl ether conversion = 100 mas. %., output mixture: 10% dimethyl ether + $N_2$, T = 450 °C, P = 0,1 MPa)

This points to an extremely complex mechanism of olefins formation at high temperatures. The observed composition of the reaction products is determined by the mechanism of the secondary reactions of changes of olefins and alkanes formed at high rates in the primary reactions.

In this case, the selectivity of propylene formation is inversely proportional to the selectivity of isobutane formation, as shown in Figure 4. It is known that a similar bond is observed in the catalytic cracking of hydrocarbons in the competition of carbon-ionic and carbon-ionic mechanisms in zeolite-containing catalysts.

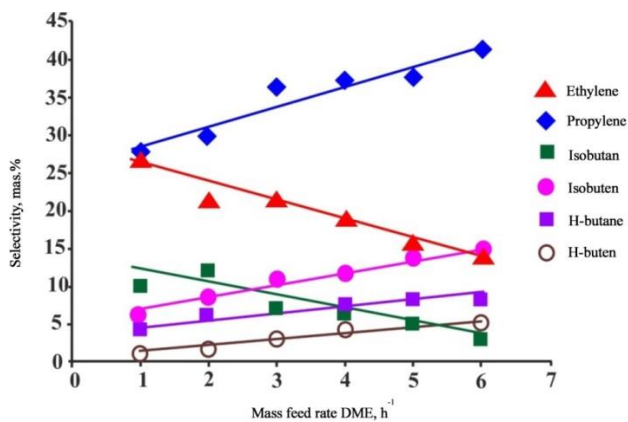

**Figure 4. The relationship between the formation of propylene and isobutane in the reaction of converting dimethyl ether to olefins.**

(Conversion of dimethyl ether = 100 mas.%., Output mixture: 10% dimethyl ether + $N_2$, T = 450 °C, P = 0,1 MPa).

As can be seen from the results presented in Table 11, in contrast to the observed laws of the reaction at 450 °C, dimethyl ether conversion in the studied range of volumetric velocities of raw material delivery decreases with decreasing contact conditional time at low temperatures, the selectivity of alkanes formation is almost independent of contact conditional time. The selectivity for the formation of butylenes and amylenes is very low and generally does not exceed 3-4 mas.%. The selectivity of ethylene and propylene formation is high (75.7-78.5 wt%) And also practically does not depend on the conversion of dimethyl ether. This indicates significant differences in the mechanism of olefin formation in primary and secondary reactions. At a relatively small rate of feed delivery (1000 h-1), the contribution of secondary reactions to the high rate of dimethyl ether is observed, and an increase in ethylene yield begins to be observed as the propylene yield decreases.

**TABLE XI. INFLUENCE OF THE VOLUMETRIC VELOCITY OF THE INITIAL GASEOUS MIXTURE ON DIMETHYL ETHER CONVERSION AND DISTRIBUTION OF REACTION PRODUCTS**

| $V_0$, h$^{-1}$ | $K_{DME}$, % | Selectivity, mas.% | | | | | | | |
|---|---|---|---|---|---|---|---|---|---|
| | | $CH_4$ | $C_{2=}$ | $C_{3=}$ | $C_{4=}$ | $C_{5=}$ | $\sum C_{2+}$ | $\sum C_{2=} - C_{4=}$ | $C_{2=}*C_{3=}$ |
| 1000 | 94,0 | 0,6 | 48,7 | 27,0 | 2,2 | 0,1 | 77,9 | 21,4 | 1,8 |
| 3000 | 57,9 | сл | 38,2 | 40,5 | 1,7 | 0,6 | 80,5 | 18,9 | 0,9 |
| 4000 | 48,2 | сл | 37,5 | 41,0 | 2,2 | 1,1 | 80,7 | 18,2 | 0,9 |
| 5000 | 32,0 | сл | 35,5 | 41,9 | 1,7 | 1,6 | 79,1 | 19,2 | 0,9 |
| 6000 | 26,1 | сл | 35,5 | 41,6 | 1,4 | 1,7 | 78,5 | 19,7 | 0,9 |
| 7000 | 28,0 | сл | 35,2 | 42,0 | 1,7 | 1,8 | 78,9 | 19,3 | 0,8 |

The magnesium-retaining catalyst is sufficiently stable after multiple oxidation regeneration cycles. As can be seen from Figure 5, the oxidative regeneration of the catalyst had almost no effect on its catalytic properties; even a slight increase in dimethyl ether conversion was observed.

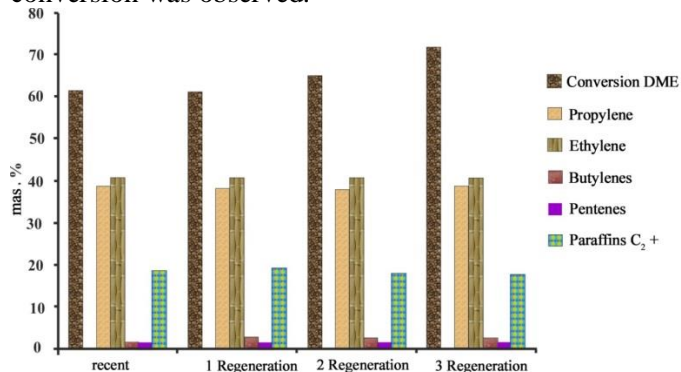

**Figure 5. Dependence of dimethyl ether conversion and selectivity on reaction products on the frequency of catalyst regeneration**

(T = 320ºC, P=0,1 MPa, output mixture: 20% dimethyl ether + N$_2$).

## VI. CONCLUSIONS

Thus, modification of the zeolite-preserving catalyst with magnesium significantly increases the selectivity for $C_2$-$C_4$ olefins (80.8 mas.%), of which almost 80 mas.% corresponds to ethylene and propylene. The volumetric speed of sending raw materials and the change in the temperature of the reaction allowed us to control a large range of the ratio of ethylene, propylene, and butene. The developed catalyst retains high activity and selectivity after oxidative regeneration.

## ACKNOWLEDGMENT

The authors acknowledge the immense help received from the scholars whose articles are cited and included in references to this manuscript. The authors are also grateful to authors/editors/publishers of all those articles, journals, and books from where the literature for this article has been reviewed and discussed.